\documentclass[aps]{JHEP3}

\usepackage{graphicx}
\usepackage{dcolumn}
\usepackage{amsfonts}
\usepackage{amsmath}
\def \be{\begin{equation}}
\def \ee{\end{equation}}
\def \bea{\begin{eqnarray}}
\def \eea{\end{eqnarray}}

\def \nn{\nonumber}
\def\a{\alpha}      
\def\b{\beta}

\def\u{\frac{u}{2}}

\def\m{\mu}

\def\G{\Gamma}

\def\bm{\begin{displaymath}}
\def\em{\end{displaymath}}

\def\sin{\mbox{sin}}
\def\cos{\mbox{cos}}

\def\m{\mbox}

\newcommand{\sigmac}{\check{\sigma}}
\newcommand{\sigc}{\sigma_{C}}
\newcommand{\sigmah}{\hat{\sigma}}
\newcommand{\nablac}{\overset{\circ}{\nabla}}
\newcommand{\chic}{\check{\chi}}
\newcommand{\chih}{\hat{\chi}}

\def\uno{\mbox{1 \kern-.59em {\rm l}}}

\newcommand{\sla}[1]{\slash\!\!\! #1}

\newcommand{\beq}[1]{\begin{eqnarray}\label{#1}}
\newcommand{\eeq}{\end{eqnarray}}

\title{On 1/2-BPS Wilson-'t Hooft loops }
\author{{Bin Chen$^1$,  and Wei He$^{2,3}$ \\
$^1$ School of Physics, Peking University ,Beijing 100871, China\\
$^2$ Institute of Theoretical Physics,Chinese Academy of
Sciences,Beijing 100080 , China\\
$^3$ Graduate University of Chinese Academy of Sciences,Beijing
100080 , China\\
E-mail: bchen01@pku.edu.cn,  weihe@itp.ac.cn }}

\abstract{ We investigate the 1/2-BPS Wilson-'t Hooft loops in
${\cal N}=4$ Super-Yang-Mills theory. We use the bulk D-brane with
both electric and magnetic charges to calculate the all genus
contribution of the circular loops. The expectation value of
Wilson-'t Hooft loops are in perfect agreement with the result
through supersymmetric condition and duality transformation in the
gauge theory.}


\begin{document}

\section{Introduction}

The AdS/CFT correspondence asserts the equivalence between type
IIB string theory in the $AdS_5\times S^5$ space with RR
background self-dual 5-form field strength and $\cal{N}$=4
super-Yang-Mills (SYM) theory\cite{Maldacena:1997, AGMOO}. It
provides a concrete example of gauge theory/string theory duality
and realizes the holographic feature of the quantum gravity in a
very special case. The AdS/CFT conjecture has passed some
nontrivial tests by performing quantitative calculation on both
side\cite{AGMOO}. One of the most impressive verification is on
Wilson loop \cite{Rey:1998ik,Maldacena:1998im}. In gauge theory,
Wilson loops can be viewed as the worldlines of a quark with the
electric charge. They define the holonomy matrices of the gauge
group along the loops. Wilson loop plays the role of the order
parameter of different phases of the gauge theory, for example it
presents the perimeter law in Higgs phase and the area law in the
confining phase. It is a natural variable to formulate the
dynamics of the gauge theory, though the resulting loop equation
is too hard to solve.

Wilson loop provides an intuitive approach to gauge theory/string
theory correspondence. Classically the loop sweep out a worldsheet
in four dimensional spacetime. If we quantize the motion of the
loop, the quantum anomaly requires another dimension: the fifth
dimension emerges holographically\cite{polya1}. If the gauge
theory is the maximally supersymmetric $\cal{N}$=4
super-Yang-Mills theory, the superconformal symmetry requires the
fifth dimension, together with $x^\mu$ to be Anti-de-Sitter space.

The $\cal{N}$=4 SYM consists of a vector multiplet
$(A_\mu,\phi^{I},\lambda_\alpha^A)$ with $\mu=1\cdots 4, I=1\cdots
6, A=1\cdots 4,\alpha=1,2$. Wilson loop in $\cal{N}$=4 SYM is
labelled by a curve $C$ in $\mathbb{R}^4$ and the representation
$R$: \be
 <W>_R(C)=\mbox{Tr}_RPe^{i\int_C
 ds(A_\mu\dot{x}^\mu(s)+\theta^I(s)\phi_I|\dot{x}(s)|)},
 \ee
 where $\theta^I(s),(I=1\cdots 6)$ is a unit vector in $\mathbb{R}^6$, defining a
 track  on $S^5$ where Wilson loop pass, $P$ denotes   the path-ordering along
 $C$. In the $AdS_5$ space the corresponding configuration is a macroscopic string
 stretched to the boundary, with the boundary condition
determined by Wilson loop. The worldsheet of the string forms a
minimal surface\cite{Drukker:1999zq} in $AdS_5$ space to minimize
its classical action. From AdS/CFT correspondence, the expectation
value of Wilson loop is just the partition function of the string
\be
 <W>_R(C)=e^{-A(C)}
 \ee
At large 't Hooft coupling the partition function can be evaluated
by the saddle-point approximation. It is the area of the minimal
surface which can be computed by Nambu-Goto action of the string.
\be
 A(C)=\int d^2\sigma \sqrt{\mbox{det}h_{ab}}
\ee As the stretched string is infinitely long, the evaluation
always suffers the divergence, which corresponds to the self
energy of the electric particle in the gauge theory. After
regulating the divergent term, we get a finite action.

For the infinitely straight line, it preserves half of the Poincare
supersymmetries. Due to non-renormalization\cite{Zarembo1,DGGM}, it
does not receive the quantum correction, its expectation value is
simply the unit \be <W>_{line}=1. \ee It means that the
corresponding worldsheet has zero regularized area.

For the circular Wilson loop, it is necessary to consider the
behavior of $\theta^I(s)$ along the loop. As analyzed in
\cite{Zarembo1}, if the direction of $\theta^I(s)$ varies along
the loop in a proper way, Wilson loop preserves half of Poincare
supersymmetries and due to non-renormalization its expectation
value is also the unit. The corresponding minimal surface in the
bulk  has zero regularized area. However, there is another kind of
the circular loop, which is obtained by a conformal transformation
of the straight line, with $\theta^I(s)$  taking constant value
along the loop. It has been explained in \cite {BGK} that this
loop preserves 16 $s$-independent linear combination of Poincare
and the conformal supersymmetries. It is called 1/2 BPS Wilson
loop. As we must choose a UV/IR cut-off in the field/string side
in the regularization process, the remaining symmetries are broken
and the expectation value receives quantum correction and depends
on the coupling constant in a non-trivial way. In
\cite{Erickson:2000af,Drukker:2000rr}, the 1/2-BPS Wilson loop in
SYM was analyzed. Based on a conformal anomaly argument, an all
order result in $1/N$ and $\lambda $ was obtained. In SYM, the
contribution to the expectation value of Wilson loop comes from
the so called rainbow graphs with the gluon propagator modified by
a singular total derivative, which give non-zero contribution only
when both ends of the propagator are located at the point which is
conformally mapped to infinity. The problem is reduced to a
quadratic hermitian matrix model with the expectation value of
Wilson loop given by
 \bea
 <W>_{circular}&=&\frac{1}{Z}\int
\mathcal{D}M\frac{1}{N}\mbox{Tr}e^Me^{-\frac{2N}{\lambda}\mbox{Tr}M^2}\nonumber\\
&=&\frac{2}{\sqrt{\lambda}}I_1(\sqrt{\lambda})+\frac{\lambda}{48N^2}
I_2(\sqrt{\lambda})+\frac{\lambda^2}{1280N^4}I_4(\sqrt{\lambda})+\cdots.
 \eea
 Where $I_n(x)$ is the first kind modified Bessel function.
In the large $N$ and large $\lambda$ limit, it simplifies to
 \be
 <W>_{circular}\sim\sqrt{\frac{2}{\pi}}\frac{e^{\sqrt{\lambda}}.}{\lambda^{3/4}}
 \sim e^{\sqrt{\lambda}}.
 \ee
In \cite{Rey:1998ik,Maldacena:1998im} the expectation value of a
circular Wilson loop is calculated by evaluating the action of the
string worldsheet, by regulating the action properly. To the
leading order in $1/N$ and $\lambda $, the result  is in agreement
with field theory calculation
 \be
 <W>_{circular}=e^{-\mbox{Area}}=e^{\sqrt{\lambda}}.
 \ee
In principle, the string worldsheet extended to the bulk could
receive higher genus contribution, this corresponds to finite N
effect in the field theory.

Recently, an important development is to calculate all genus
contributions to Wilson loop using bulk D-brane in the string theory
side. It is first proposed in \cite{Rey:1998ik} for the fundamental
representation and followed by
\cite{Fiol,Hartnoll:2006hr,GP,Yamaguchi:2006D5} for higher rank
symmetric and antisymmetric representations. The reason is that the
string worldsheet in the five form field strength background can
blow up in the transverse direction to form the dielectric
brane\cite{Myers:1999,Pawelczyk:2000hy,Rodriguez2006,Callan:1997},
analogous to the formation of the gaint
gravitons\cite{Grisaru:2000zn,Hashimoto:2000zp,McGreevy:2000cw}. Now
the expectation value of Wilson loop is the partition function of
D-brane with string charges on it. It is obtained by evaluating the
D-brane action instead of the string action. On the gauge theory
side, the corresponding quantities are also evaluated in the matrix
model\cite{Hartnoll:2006hr,Akeman,OS,OGS}.

It is analyzed in \cite{GP} that the string worldsheet can
blows up in two ways, producing two kinds of branes:\\
$\bullet$ The strings blow up in $S^2\subset AdS_5$, resulting in
a D3 brane with induced metric $AdS_2\times S^2$ and $k$ units of
fundamental string charge. This
corresponds to Wilson loop in the symmetric representation of rank $k$.\\
$\bullet$ The strings blow up in $S^4\subset S^5$, resulting in a D5
brane with induced metric $AdS_2\times S^4$ and $k$ unites of
fundamental string charge. This corresponds to Wilson loop in
anti-symmetric representation of rank $k$.

For Wilson loop in the symmetric representation, both D-brane
calculation and the matrix model calculation yield the following
result
 \be
  <W>_{\mbox{sym}}=e^{2N(\kappa\sqrt{1+\kappa^2}+\sinh^{-1}\kappa)}.
 \ee
with $\kappa=\frac{k\sqrt{\lambda}}{4N}$. For Wilson loop in the
anti-symmetric representation, D-brane calculation and matrix
model calculation yield
 \be
 <W>_{\mbox{asym}}=e^{\frac{2N\sqrt{\lambda}}{3\pi}\sin^3\theta_k}.
 \ee
with $\theta_k-\frac{1}{2}\sin 2\theta_k=\frac{\pi k}{N}$.

In the gauge theory there is another nonlocal operator: 't Hooft
loop, which is the magnetic dual of Wilson loop. It is defined as
a singular gauge transformation along the loop and can be viewed
as the worldline of a particle with magnetic charge. Its
expectation value behaves oppositely to Wilson loop: the perimeter
law in the confining phase and the area law in Higgs phase. In a
general gauge theory, the property of 't Hooft loop is hard to
calculate because the electric-magnetic duality is rather obscure.
In the $\cal{N}$=4 SYM, which inherits a $\mbox{SL}(2,Z)$ duality
and is self-dual, it is quite natural to consider 't Hooft loop
together with Wilson loop. For simplicity we set the axion to be
zero first and discuss the full $\mbox{SL}(2,Z)$ duality later.
Under S-duality, Wilson loop becomes 't Hooft loop, the
fundamental string (F1) becomes D-string (D1), so the macroscopic
object corresponding to 't Hooft loop is the D-string worldsheet.
Analogous to Wilson loop case we can use D3/NS5 brane with
D-string charge  to calculate the all genus contribution.

In this paper,  we will study 1/2 BPS Wilson-'t Hooft loops. From
the gauge theory point of view, Wilson-'t Hooft loop is the
worldline of a dyon which carry both the electric charge and
magnetic charge\cite{Kapustin2005}. From string theory point of
view, Wilson-'t Hooft loops are the (F1, D1) bound
states\cite{WITTENboun} with the strings ending on the worldvolume
of the D3 branes. The charge of (F1, D1) corresponds to the charge
of the dyon. However there is a subtlety. Wilson loop can be in any
representation of the gauge group. For the symmetric representation
of rank $k$, the corresponding fundamental string is a single string
with charge $k$, while for Wilson loop in the antisymmetric
representation, the corresponding fundamental strings are $k$ F1's
each with charge one. For a generic representation, the
configuration of the fundamental strings is quite complicated. For
Wilson-'t Hooft loops, the existence of D-string make things more
subtle. Similar to the F1's, the D1's could have different kinds of
combination, corresponding to the different representation of the
dual group. In this paper, we will consider the case that both the
F1's and D1's are in the symmetric representation. In other words,
the F1's and D1's form a simple bound state.

The paper is organized as follows. In section 2, we will calculate
the all-genus contribution to 1/2 BPS Wilson-'t Hooft loops using
bulk D3-brane. We consider straight Wilson-'t Hooft line first and
then the circular Wilson-'t Hooft loop. In both cases, the
calculations shows that the expectation value of the line/loop
depends on the charges and the coupling constant in a S-dual
invariant way. In section 3, we analyze the supersymmetry of the
bulk D3-brane with fluxes, it preserves half of the original
supersymmetries. In section 4, we discuss WH loop in the
antisymmetric representation related to bulk 5-brane
configurations. We end the paper with conclusion and discussions.

\section{All-genus calculation using D3-branes}

In  \cite{Fiol}, it has been shown that one can use the bulk
D3-brane to evaluate Wilson loop operators. By evaluating the
classical action of D3 brane with $n$ F-string charge, the all
genus expectation value of the circular BPS Wilson loop is
 \be
 <W>=e^{2N(\kappa\sqrt{1+\kappa^2}+\sinh^{-1}\kappa)}
 \ee
with $\kappa=\frac{n\sqrt{\lambda}}{4N}$ and $n$ is the F-string
charge on D3-brane. The all genus expectation value of the
circular BPS 't Hooft loop is also evaluated
 \be
 <H>=e^{2N(\kappa\sqrt{1+\kappa^2}+\sinh^{-1}\kappa)}
 \ee
now with $\kappa=\frac{\pi m}{\sqrt{\tilde{\lambda}}}$ and $m$ is
the D-string charge on D3-brane. Here
$\tilde{\lambda}=\frac{16\pi^2N^2}{\lambda}$ is the dual coupling
constant.

 In this section we will evaluate all
genus expectation value of the straight and the circular BPS
Wilson-'t Hooft (WH) loop using D3 brane with both F-string and
D-string charges. The F-string and D-string would form bound state
preserving $1/2$ supersymmetry\cite{WITTENboun}. For a
$(n,m)$-string, the string tension is
 \be
 \tau_{n,m}=\frac{\sqrt{n^2+m^2/g_s^2}}{2\pi\alpha^{'}}
 \ee
which is invariant under S-duality:
$(n,m,g_s,\alpha^{'})\leftrightarrow (-m,n,g_s^{-1},\alpha^{'}g_s)$.
In terms of field theory coupling constant $ \lambda$ it is
 \be
 \tau_{n,m}=\frac{1}{\pi\alpha^{'}\sqrt{\lambda}}.
 \sqrt{n^2\lambda+m^2\tilde{\lambda}}
 \ee
  When the $(n,m)$-string ends on the D3-brane, the
resulting system is still BPS. We expect that the expectation
value takes the same form as Wilson loop, with a suitable
dependence on $\sqrt{n^2\lambda+m^2\tilde{\lambda}}$. Our
calculation shows it is indeed the case.

In the calculation an essential point is to take into account of
the boundary contribution to the action. As D3 brane in the bulk
intersects boundary along the loop, we should take the  possible
boundary terms into account. The boundary terms implement the
Legendre transformations to let the solution have the correct
boundary conditions. There are two kinds of the boundary terms:
one comes from the Neumann boundary condition  on the transverse
radial direction; the other comes from the Legendre transform of
the gauge fields such that the charges on the brane worldvolume is
fixed.

For coordinates $x^i$ satisfying Neumann boundary condition, we
need to perform a Legendre transformation on the brane action to
make the action to be a function of $p_i$:
 \be
\tilde{S}=S-\int_C ds p_ix^i,
 \ee
where $p_i$ is the conjugate momentum of $x^i$
  \be
p_i=\frac{\partial S}{\partial \dot{x^i}}.
 \ee
For $AdS_5\times S^5$ space, the coordinates take Dirichlet boundary
condition on the direction parallel to the boundary of $AdS$ and
take Neumann boundary condition on the remaining directions. As D3
brane does not extend to $S^5$ part, we should take into account
only the boundary term related to the radial direction.

Similarly, for the gauge field $A_i$ on the brane, its conjugate
momentum is
  \be
\Pi_i=\frac{\partial S}{\partial \dot{A_i}}
 \ee
the corresponding boundary term is
 \be
-i\int_C ds \Pi_iA_i.
 \ee
Here we include $i=\sqrt{-1}$ factor because  when evaluating the
action we always use Euclidean version of $AdS_5\times S^5$
metric. In the previous work evaluating Wilson loop using the
string worldsheet, there is no boundary term related to gauge
field.

\subsection{Straight Wilson-'t Hooft line}

Let us start from the infinite straight Wilson-'t Hooft line in
$\mathbb{R}^4$. It is convenient to use the following coordinate
system for $AdS_5$:
 \be
 ds^2=\frac{L^2}{y^2}(dy^2+(dx^1)^2+dr^2+r^2(d\theta^2+\sin^2\theta
 d\varphi^2)).
 \ee
 Where $L^4=\lambda\alpha^{'2}$ is the radius
 of $AdS_5$ and $S^5$.
 The straight Wilson-'t Hooft line lies along $X^1$ and is
 localized in the transverse directions. One may understand the
 physics from dielectric effect and try to calculate the
 expectation value of Wilson-'t Hooft line from the D3-brane action. As has
 been shown in \cite{Fiol}, the worldvolume of D3-brane is a
 hypersurface in $AdS_5$, which is characterize by the equation
 $y=y(r)$ and worldvolume coordinates $(x^1,r,\theta,\varphi)$.

 For $D3$ brane ending on the boundary of $AdS_5$, the action
 includes three parts: the Dirac-Born-Infeld action, the Wess-Zumino
 action and the boundary term
 \be
S_{t}=S_{DBI}+S_{WZ}+S_{boundary}
 \ee
 The Dirac-Born-Infeld action is of the form
  \be
  S_{DBI}=-T_{D3}\int e^{-\Phi} \big(\frac{L^2}{y^2}\big)^2\sqrt{\big(1+y^{\prime
  2}+(\frac{y^2}{L^2})^2(2\pi\a^\prime)^2F^2_{tr}\big)\big(r^4\sin^2\theta+(\frac{r^2}{L^2})^2
  (2\pi\a^\prime)^2F^2_{\theta\varphi}\big)}
  \ee
  where $F_{tr}$ is the electric field and $F_{\theta\varphi}$ is
  the magnetic field. And the tension of the D3-brane is
  \be
  T_{D3}=\frac{N}{2\pi^2L^4}.
  \ee
The Wess-Zumino term contributes to the action
 \be
  S_{WZ}=\mu_{D3}\int P[C_4]=-\frac{2N}{\pi}\int
  dx^1dr\frac{r^2}{y^4}
  \ee
  where $\mu_{D3}=T_{D3}$ is the charge of D3 brane, $P[C_4]$ is the pullback of the Ramond-Ramond 4-form
  potential to the worldvolume of D3-brane and
   \be
   C_4=\frac{L^4r^2\sin\theta}{y^4}dx^1\wedge dr \wedge d\theta
   \wedge d\varphi
   \ee
   The total bulk action is just $S_{bulk}=S_{DBI}+S_{WZ}$.

 Without the magnetic field, we have simply the straight Wilson
 line and the electric field takes the form
  \be\label{E}
  F_{1r}=i\frac{n\lambda}{8\pi Nr^2}
  \ee
  which solves the equation of motion and spread over the worldvolume of
  $AdS_2$ uniformly. Here the appearance of
  $i$ is also due to Euclidean metric. In this
  case, it has been shown that the action of D3-brane is vanishing
  due to the cancellation between DBI and WZ action. Even after
  considering the boundary terms, the action is still vanishing,
  implying that the expectation value of Wilson line $<W>=1$ which has
  to  be true due to the supersymmetric condition. In the case
  without the electric field, we have the straight 't Hooft line,
  which is the S-dual of Wilson line. The corresponding magnetic
  field takes the   form
  \be\label{M}
  F_{\theta\varphi}=\frac{m\sin\theta}{2}
  \ee
which spread over the $S^2$ uniformly. The straight 't Hooft line
has not been treated carefully in the literature but due to the
supersymmetry and S-duality, it has been expected that $<H>=1$. We
will not discuss it separately. Instead, we will discuss the more
general case with both electric and magnetic charges, namely
Wilson-'t Hooft line.

In the case of Wilson-'t Hooft line, the electric field and
magnetic field take the form as (\ref{E},\ref{M}). And we make the
linear ansatz \be y=\frac{r}{\kappa}. \ee With these setups, the
bulk action is not vanishing and instead diverges near $y=0$. Let
us introduce a cutoff $y_0$ and the bulk action is
 \be
 S_{bulk}=-\frac{2N}{\pi}X^1 (bc-1)\frac{\kappa^3}{y_0}
 \ee
 where
 \bea
 b&=&\sqrt{1+\kappa^{-2}-\frac{n^2\lambda}{16N^2}\kappa^{-4}} \label{b}\\
 c&=&\sqrt{1+(\frac{\pi\a^\prime m}{L^2})^2} \label{c}
 \eea
 On the other hand, one has to take into account of the boundary
 terms\cite{Fiol}. One kind of such boundary terms is
 \be
 -\int dx^1 y_0p_y
 \ee
 where $p_y$ is the conjugate momentum to $y$ after integration over $S^2$:
  \be
  p_y=-\frac{2N}{\pi}\frac{\kappa}{y^2_0}\frac{c}{b}.
  \ee
  The other boundary term comes from the Legendre transform of the
  gauge field:
  \be
  -\int dx^1(i\Pi)A_1=-\int
  dx^1dr(i\Pi)F_{1r}=-\frac{c}{b}X^1\frac{n^2\lambda}{8\pi
  N}\frac{\kappa^{-1}}{y_0}.
  \ee
After putting all the contribution together, the total action is
 \be
 S_t=-\frac{2\pi}{N}X^1 Z\frac{1}{y_0}
 \ee
 where
 \be\label{Z}
 Z=\kappa^3(bc-1)+\frac{c}{b}(\frac{n^2\lambda}{16N^2}\kappa^{-1}-\kappa).
 \ee
The divergence near $y_0\rightarrow 0$ should be absent, this leads
to the condition $Z=0$. This helps us to fix the value of $\kappa$:
\be\label{kappa2}
\kappa^2=\frac{n^2\lambda}{(4N)^2}+\frac{\pi^2m^2}{\lambda}. \ee
Using the dual coupling constant
$\tilde{\lambda}=\frac{16\pi^2N^2}{\lambda}$, $\kappa$ can be
written as \be\label{kappa2}
\kappa^2=\frac{n^2\lambda}{(4N)^2}+\frac{m^2\tilde{\lambda}}{(4N)^2}
\ee

With the divergence vanishing, we have $S_{\mbox{tot}}=0$ so that
$<WH>=1$. This is consistent with the BPS condition.

\subsection{Circular Wilson-'t Hooft loop}

For the circular Wilson loop, it could be related to the straight
Wilson line by a conformal transformation. Due to the quantum
anomaly of the conformal transformation, the expectation value of
the circular Wilson loop is not $1$. On the N=4 SYM side, the
calculation reduces to a quadratic Hermitian matrix model
\cite{Erickson:2000af,Drukker:2000rr}. For Wilson loop in the rank
$k$ symmetric representation, $<W>_S=<\frac{1}{N}Tr e^{nM}>$. The
calculation in the Matrix model shows that to all orders in
$g_{YM}$ and $\frac{1}{N}$ expansion:
 \be \label{Ws}
 <W>_S=e^{2N[\kappa\sqrt{1+\kappa^2}+\sinh^{-1}\kappa]}
 \ee
 which is in perfect agreement with the calculation from bulk D3-brane
 on the dual string theory side. Furthermore, the expectation value of 't Hooft loop
 in the symmetric representation of the dual group has also been
 worked out in \cite{Fiol}. It takes the same form of (\ref{Ws}),
 but with a different $\kappa$.

 For Wilson-'t Hooft circular loop, we don't know how to
 calculate it from Super-Yang-Mills theory since we have to work
 in a background with magnetic monopole. Nevertheless, the BPS
 condition and the $SL(2, Z)$ duality in the ${\cal N}=4$ SYM
 permits us to get the answer. Actually, in both Wilson and 't Hooft cases
 $\kappa$'s
 are proportional to the tension of the corresponding macroscopic strings. This fact leads us
 to predict that for Wilson-'t Hooft loop, the expectation value
 should take the same form as (\ref{Ws}), but with the $\kappa$
 being proportional to  the tension of (F1, D1) bound state.

Consider D3 brane with $n$ F1 string charge and $m$ D1 brane charge,
The DBI action of D3 brane is
\be S_{DBI}=-T_{D3}\int
e^{-\Phi}\sqrt{det(g+2\pi\alpha^{'}F)} \ee
 To describe the circular
loop, the convenient metric of $AdS_5$ space is
\be
ds^2=\frac{L^2}{\sin^2\eta}(d\eta^2+\cos^2\eta
d\psi^2+d\rho^2+\sinh^2\rho(d\theta^2+\sin^2\theta d\phi^2)) \ee
 In
these coordinates, $\rho\in [0,\infty), \theta\in[0,\pi]$ and
$\eta\in [0,\pi/2]$. The boundary of the AdS space is now at
$\eta\rightarrow 0, \rho\rightarrow 0$ and the circle on the
boundary is located at $\eta=\rho=0$. According  to the symmetry,
we take $\rho,\psi,\theta,\phi$ as the coordinates of D3 brane
worldvolume which is curved in $(\eta,\rho)$ plane according to
$\eta=\eta(\rho)$.

 Take the following ansatz for the gauge field
\be F=F_{\psi\rho}d\psi\wedge d\rho+F_{\theta\phi}d\theta\wedge
d\phi \ee The DBI action is \be S_{DBI}=-\int d\rho d \theta
2N\frac{\sinh^2\rho\sin\theta}{\sin^4\eta}\sqrt{\big(\cos^2\eta(1+\eta^{'2})+(2\pi
\alpha^{'})^2\frac{F_{\psi\rho}^2\sin^4\eta}{L^4}\big)\big(1+(2\pi\alpha^{'})^2\frac{F_{\theta\phi}^2
\sin^4\eta}{L^4\sin^2\theta\sinh^4\rho}\big)} \ee The WZ action is
\be S_{WZ}=2N \int d\rho d \theta
\frac{\cos\eta\sin\theta\sinh^2\rho}{\sin^4\eta}(\cos\eta+\eta^{'}\sin\eta
\frac{\sinh\rho-\cosh\rho\cos\theta}{\cosh\rho-\sinh\rho\cos\theta})
\ee The equations of motion are solved by setting
\be\label{etakappa} \sin\eta=\kappa^{-1}\sinh\rho \ee with $\kappa$
a function of charges and coupling constant \be\label{cirkappa2}
\kappa^2=\frac{n^2\lambda}{(4N)^2}+\frac{m^2\tilde{\lambda}}{(4N)^2}\ee
And the gauge fields take value \be\label{cirEB}
F_{\psi\rho}=\frac{in\lambda}{8\pi N \sinh^2\rho} \qquad
F_{\theta\phi}=\frac{m\sin\theta}{2} \ee
 These are just the same
electric and magnetic fields (\ref{E},\ref{M}) in the new
coordinates.

The DBI action becomes
\be
S_{DBI}=-2N \int d\rho d \theta
\frac{\sinh^2\rho\sin\theta}{\sin^4\eta}bc
\ee
where $b$ and $c$ are defined by
\begin{eqnarray}\label{bc}
b&=&\sqrt{\cos^2\eta(1+
\eta^{'2})-\frac{n^2\lambda\sin^4\eta}{(4N)^2\sinh^4\rho}}\nonumber\\
c&=&\sqrt{1+\frac{m^2\tilde{\lambda}\sin^4\eta}{(4N)^2\sinh^4\rho}}
\end{eqnarray}

Then the conjugate momentum of $\eta$ is
\begin{eqnarray}
p_{\eta}&=&\frac{\delta S}{\delta
\eta^{'}}=-2N\frac{\kappa^2\sin\theta\cos^2\eta\eta^{'}}{\sin^2\eta}
\frac{c}{b}\nonumber\\
\quad &+&2N \frac{\kappa^2\cos\eta\sin\theta}{\sin\eta}
(\frac{\sinh\rho-\cosh\rho\cos\theta}{\cosh\rho-\sinh\rho\cos\theta})
\end{eqnarray}
The boundary term for $p_{\eta}$ at $\eta_0\to 0$ is \be
S_{p_{\eta}}=-\eta_0\int p_{\eta}=-4N\kappa
\frac{1}{\eta_0}\frac{c}{b} \ee The boundary term for gauge field is
\begin{eqnarray}
S_{F}&=&-i\int d\rho d\psi \Pi F_{\psi\rho}\nonumber\\
\quad &=&-
\frac{n^2\lambda}{4N}\frac{c}{b}\mbox{coth}\rho|_{\sinh\rho=\kappa\sin\eta_0}
^{\sinh\rho=\kappa}
\end{eqnarray}
where $\Pi$ is the conjugate of $F_{\psi\rho}$. The total action is
\begin{eqnarray}
S_{D3}&=&-4N\kappa^4(bc-1)\mbox{coth}\rho|_{\sinh\rho=\kappa\sin\eta_0}
^{\sinh\rho=\kappa}+
2N\kappa^2(\mbox{coth}\rho-\frac{\rho}{\sinh^2\rho})|_{\sinh\rho=\kappa\sin\eta_0}^{\sinh\rho=\kappa}\nonumber\\
\quad &-& 4N\kappa \frac{1}{\eta_0}\frac{c}{b}-
\frac{n^2\lambda}{4N}\frac{c}{b}\mbox{coth}\rho|_{\sinh\rho=\kappa\sin\eta_0}
^{\sinh\rho=\kappa}
\end{eqnarray}
where $b,c$ are the same as (\ref{bc}). With $\kappa$ in
(\ref{cirkappa2}), divergent terms when $\eta_0\to 0$ cancel,
 we have once again \be
Z=\kappa^3(bc-1)+\frac{c}{b}(\frac{n^2\lambda}{16N^2}\kappa^{-1}-\kappa)=0
\ee  Then the action is \be
S_{D3}=-2N(\kappa\sqrt{1+\kappa^2}+\mbox{sinh}^{-1}\kappa), \ee
and the expectation value of Wilson-'t Hooft loop is \be
<WH>=e^{-S_{D3}}=e^{2N(\kappa\sqrt{1+\kappa^2}+\mbox{sinh}^{-1}\kappa)}
\ee Therefore, we have shown that even for Wilson-'t Hooft
circular loop, the bulk D3-brane action gives the all-genus
contribution which is consistent with the field theory argument
based on duality.

In the limit $N\to \infty$, the expectation value of WH loop
simplifies to \be <WH>_{N\to
\infty}=e^{\sqrt{n^2\lambda+m^2\tilde{\lambda}}} \ee which is the
same as the prediction of \cite{BGK} based on duality. The reason is
that
 the F-strings and D-strings form bound state on D3 brane,
 resulting in a
$(n,m)$ string. The relation (\ref{kappa2}) could be rewritten in
another way:
 \be
 \kappa^2=\frac{\pi g_s}{4N}(n^2+\frac{m^2}{g^2_s}).
 \ee
The term in the bracket is proportional to the square of the
tension of the (F1, D1) bound states with charge $(n,m)$. This is
natural since in the brane picture the  dyon in the
super-Yang-Mills theory corresponds to the (F1,D1) bound state
ending on the D3-brane. In the super-Yang-Mills, the dyon
solutions are BPS, their energy-charge relation satisfy the BPS
relation:
 \be
M=\sqrt{2}|Z|=<\phi>\sqrt{n^2g_{YM}^2+\frac{m^2}{g_{YM}^2}},
 \ee
which is invariant under S-duality,$<\phi>$ is the vacuum expectation
value of the scalar field. So the dependence on charges and couplings
matches on both sides.

 The induced metric on D3 brane is
\be g_{ab}=\partial_a X^{\mu}\partial_b X^{\nu}G_{\mu\nu} \ee The
only non-trivial component is \be
g_{\eta\eta}=\frac{L^2}{\sinh^2\eta}(\frac{1+\kappa^2}{1+\kappa^2\sin^2\eta})
\ee So the induced metric is \be
ds_{D3}^2=\frac{L^2}{\sinh^2\eta}(\frac{1+\kappa^2}{1+\kappa^2\sin^2\eta})d\eta^2
+\frac{L^2\cos^2\eta}{\sin^2\eta}d\psi^2+L^2\kappa^2(d\theta^2+\sin^2\theta
d\phi^2) \ee Change variable
$\mbox{cot}^2\eta=(1+\kappa^2)\sinh^2\zeta$, the metric is
Euclidean $AdS_2\times S^2$ with different radius: \be
ds_{D3}^2=L^2(1+\kappa^2)(\sinh^2\zeta d
\psi^2+d\xi^2)+L^2\kappa^2(d\theta^2+\sin^2\theta d\phi^2) \ee The
radius of $AdS_2$ is $L\sqrt{1+\kappa^2}$ while the radius of
$S^2$ is $L\kappa$. Embedding into $AdS_5\times S^5$, we can write
the $AdS_5\times S^5$ metric as the fibration of 2-dimensional
space with $AdS_2\times S^2 \times S^4$ fiber by defining two new
variables $u,\zeta$: \be \mbox{cot}\eta=\cosh u\sinh\zeta \qquad
\sinh\rho=\sinh u\sin\eta \ee then the metric of $AdS_5\times S^5$
is of the form
\be \label{fibre} ds^2=L^2\lbrack\cosh^2u
d\Omega_{AdS_2}^2+du^2+d\theta^2+\sinh^2ud\Omega_{S^2}^2
+\sin^2\theta d\Omega_{S^4}^2 \rbrack \ee with $\sinh u=\kappa$.

In the new fibred coordinates, the gauge field strength reads as
 \be
 F=-\frac{\sqrt{1+\kappa^2}}{\kappa}(\frac{in\lambda}{8\pi
 N})\sinh\zeta d\psi\wedge d\zeta+\frac{m\sin\theta}{2}d\theta\wedge
 d\phi
 \ee

\subsection{Restore the full $\mbox{SL}(2,Z)$ duality}

In the previous sections we don't take the theta dependence into
account and set the axion field to zero in the D-brane
calculation. In this subsection we will consider the effect of a
nonvanishing constant axion field  and try to restore the full
$\mbox{SL}(2,Z)$ duality of the theory\cite{Kapustin2005}.

For a non-zero constant axion background, it contributes the
following term to the Wess-Zumino action \be
S_{\m{axion}}=i\mu_3(2\pi\alpha)^2\int
C_0F_{\psi\rho}F_{\theta\phi}=i2\pi C_0\int d\rho d\theta
F_{\psi\rho}F_{\theta\phi} \ee where $i$ is due to the  Euclidean
signature.

The bulk action now is \be S_{\m{bulk}}=S_0+S_{\m{axion}}. \ee The
equations of motion and the charge quantization conditions for
F/D-string charges are solved by \be\label{etakappaaxion}
\sin\eta=\kappa^{-1}\sinh\rho \ee with $\kappa$ taking value
\be\label{cirkappaaxion}
\kappa^2=\frac{(n+mC_0)^2\lambda}{(4N)^2}+\frac{m^2\tilde{\lambda}}{(4N)^2}
\ee and the gauge fields taking values \be\label{cirEBaxion}
F_{\psi\rho}=\frac{i(n+mC_0)\lambda}{8\pi N \sinh^2\rho} \qquad
F_{\theta\phi}=\frac{m\sin\theta}{2} \ee All the calculations are
in parallel with the ones in the zero axion case with a
replacement $n\to n+mC_0$.

From the AdS/CFT correspondence, the axion in the bulk could be
identified with the $\theta$ parameter in the Yang-Mills theory
\be C_0=\frac{\theta}{2\pi}, \ee so the $\mbox{SL}(2,Z)$ invariant
result is \be
<W>_{\m{sym}}=e^{2N(\kappa\sqrt{1+\kappa^2}+\mbox{sinh}^{-1}\kappa)}
\ee with
\begin{eqnarray}
\kappa^2&=&(n+\frac{m\theta}{2\pi})^2\frac{\lambda}{(4N)^2}+
\frac{m^2\tilde{\lambda}}{(4N)^2}\nn\\
&=&\frac{\pi}{4N}\frac{|n+m\tau|^2}{\mbox{Im}\tau}
\end{eqnarray}
which is invariant under the S and T transformation \bea
S&:&\tau\to -\frac{1}{\tau} \qquad (n,m)\to (-m,n)\\
T&:&\tau\to\tau+1 \qquad (n,m)\to(n+m,m) \eea

This is what we expect from both the field theory and the dual
string theory: on the field theory side, in the $\theta$-vacuum
the magnetic charge will induce a fractional electric charge. The
shift is exactly $\frac{m\theta}{2\pi}$; on the dual string theory
side, when the axion field is non-zero, the D-string charge will
induce a F-string charge $mC_0$.

\section{Supersymmetry of $AdS_2\times S^2$ D3-brane with fluxes}

In this section, we would like to discuss the preserved
supersymmetry of the $AdS_2\times S^2$ D3-brane with fluxes. Now we
change to Minkowski metric and multiply the gauge field strength on
the $AdS_2$ part  by $i$. The Gamma matrices and Killing spinor
convention will follow \cite{Yamaguchi:2006te,Yamaguchi:2006D5}. To
be consistent, we put them in the appendix A.

To discuss the supersymmetry of the D-branes in a curved
background, one has to check the kappa symmetry projection
relation \be \G_{p}\epsilon=\epsilon \ee where $\G_{p}$ is
determined by the embedded curved spacetime and background
fluxes\cite{Bergshoeff:1996,Bergshoeff:1997,Skenderis:2002vf,Imamura:1998gk}.
For a Dp-brane in IIB theory, it is determined by the following
relations:
 \bea
 \G_{Dp} d^{p+1}\sigma&=&-e^{-\Phi}(-det(G+{\cal F}))^{-\frac{1}{2}}e^{\cal
 F}\wedge\chi|_{\mbox{$(p+1)$-form}} \nn\\
 \chi&=&\sum_n\frac{1}{(2n)!}d\sigma^{i_2n}\wedge\cdots\wedge
 d\sigma^{i_1}\G_{\langle i_1i_2\cdots
 i_{2n}\rangle}\tau_3^n(i\tau_2) \nn\\
 \G_{\langle i_1i_2\cdots
 i_{s}\rangle}&=&\frac{\partial X^{\mu_1}}{\partial \sigma_{i_1}}\cdots
 \frac{\partial X^{\mu_s}}{\partial
 \sigma_{i_s}}e^{a_1}_{\mu_1}\cdots e^{a_s}_{\mu_s}\G_{a_1\cdots
 a_s}
 \eea
 where $\sigma^i$ are the worldvolume coordinates, $X^\mu$ are
 the spacetime coordinates and $e^a_\mu$ are the components of the
 vielbein. And as usual ${\cal F}=B+2\pi\alpha^\prime F$. With
 respect to the metric (\ref{fibre}), the vielbein relevant to our discussions are
 \be
 e^0=L\cosh u f^0, \hspace{3ex} e^1=L\cosh u f^1,
 \hspace{3ex}e^4=L\sinh u f^4, \hspace{3ex} e^5=L\sinh u f^5 \ee
 where $f^0, f^1$ are the vielbein of the unit $AdS_2$ and $f^4,
 f^5$ are the vielbein of the unit $S^2$.

In the D3-brane picture, D3-brane worldvolume extends along
$AdS_2\times S^2$ directions and sits at the point on $S^5$ with $\theta=0$
or $\pi$.  In our case, we turn on the electric and magnetic flux
at the  same time. The electric field spread over $AdS_2$
 and the magnetic field over $S^2$ uniformly such that
  \be
  {\cal F}=\alpha e^0_0e^1_1d\sigma^0\wedge d\sigma^1+\beta
  e^4_4e^5_5d\sigma^4\wedge d\sigma^5,
  \ee
where $(\sigma^0,\sigma^1,\sigma^4,\sigma^5)$ are  the worldvolume
coordinates.
 Then we have
 \be
 \G_{D3}=-\frac{1}{A}((i\a\b-\mu_3\nu_3)\tau_2+(i\a\nu_3+\b\mu_3)\tau_1)
 \ee
 with
 \be
 A=\sqrt{(1-\a^2)(1+\b^2)}
 \ee

 The projection condition reads
 \be
 \G_{D3}\epsilon=\epsilon
 \ee
 with
 \be
 \epsilon=\exp(\frac{1}{2}\mu_3\nu_3\tau_2
 u-\frac{i}{2}\lambda_3\tau_2\theta)\zeta
 \ee

 Let's first focus on the case with $\theta=0$. When we have only
 electric flux, i.e. $\a\neq 0, \b=0$, the situation reduce to the BPS Wilson loop and
 has been analyzed in \cite{Yamaguchi:2006D5}:
  \be
  \sinh u_k=\frac{\sqrt{1-\a^2}}{|\a|}\hspace{3ex}, \left\{\begin{array}{ll}
  (1-\mu_3\tau_3)\zeta=0,& \a>0 \\
  (1+\mu_3\tau_3)\zeta=0,& \a<0
  \end{array} \right.
  \ee
  On the other hand, when
 $\a=0, \b\neq 0$, the supersymmetric condition has not been
 discussed in the literature. Actually, the projection condition
 leads to
\be
  \sinh u_k=\frac{1}{|\b|}\hspace{3ex}, \left\{\begin{array}{ll}
  (1-\mu_3\tau_1)\zeta=0,& \b<0 \\
  (1+\mu_3\tau_1)\zeta=0,& \b>0
  \end{array} \right.
  \ee
Let us turn on the electric and magnetic flux at the same time,
i.e. $\a\neq 0, \b \neq 0$. The supersymmetric condition is
\begin{align}\label{SUSY1}
\left\{(\cosh\u+\frac{\sinh\u(i\alpha\beta\mu_3\nu_3-1)}{A})
+(\mu_3\nu_3\sinh\u+\frac{\cosh\u(i\alpha\beta-\mu_3\nu_3)}{A})\tau_2\right.\nonumber\\
\left. +\frac{\cosh\u(i\alpha\nu_3+\beta\mu_3)}{A}\tau_1
+\frac{\sinh\u(-\alpha\mu_3+i\beta\nu_3)}{A}\tau_3\right\}\zeta=0.
\end{align}

From the discussion above, we make the following ansatz on
$\zeta$:
 \be
 (1-(x\mu_3\tau_3+y\mu_3\tau_1))\zeta=0,
 \ee
where $x,y$ are two real numbers satisfying $x^2+y^2=1$. Then the
supersymmetric condition leads to the following solutions:
\begin{align}\label{xy}
x=\alpha\sqrt{\frac{1+\beta^2}{\alpha^2+\beta^2}} \qquad
y=-\beta\sqrt{\frac{1-\alpha^2}{\alpha^2+\beta^2}}
\end{align}
and
 \be\label{uk}
  \sinh u_k=\sqrt{\frac{1-\a^2}{\a^2+\b^2}}.
  \ee
It is still one step from proving that the configurations
discussed in section 2 actually is 1/2-BPS. Let us identify the
gauge field strength parameters
\bea
\a&=&\frac{n\sqrt{\lambda}}{4N\kappa\sqrt{1+\kappa^2}} \nn\\
\b&=&\frac{m\pi}{\sqrt{\lambda}\kappa^2}=\frac{m\sqrt{\tilde{\lambda}}}{4N\kappa^2}
\eea and put them back to the relation (\ref{uk}). The relation
holds automatically, taking into account of the relation
(\ref{cirkappa2}). This indicates that the D3-brane with electric
and magnetic field (\ref{cirEB}) satisfy the 1/2-BPS configuration
if its motion parameter satisfy (\ref{etakappa}).

For the case of $\theta=\pi$, we have very similar result. In
either case, the configuration keep one half of the original
supersymmetry.

For nonvanishing axion background, the bulk supersymmetry analysis
here still holds because the kappa projection operator dosen't
involve any RR field. The solution of the supersymmetry condition
(\ref{xy},\ref{uk}) with $n\rightarrow n+mC_0$ is satisfied by
(\ref{cirEBaxion}). In the dual SYM, the $\theta$-term contributes
an extra term to the central charge of the SUSY algebra, the
solution (\ref{cirEBaxion}) still saturates the BPS condition.

\section{The calculation using 5-brane}

It was shown in \cite{Yamaguchi:2006D5} the expectation value of
Wilson loop in anti-symmetric representation could be calculated
using bulk D5-brane action. Remarkably, the calculations from
D5-brane and the Matrix theory are in perfect match. It is
interesting to investigate if  Wilson-'t Hooft loop in
anti-symmetric representation has the same story.

Let us study 't Hooft loop first. In IIB theory, the S-dual of D5
brane is NS5 brane. Hence 't Hooft loop in antisymmetric
representation  corresponds to NS5 brane with D-string charges.
But now from NS5-brane point of view, the D-string looks like
electric source rather than magnetic source. This is very
different from the D3-brane case. The fields on IIB NS5 brane is a
$(1,1)$ vector
 supermultiplet, consisting of four scalars and a vector one form $c^{(1)}$.
 The effective worldvolume action of
NS5 brane was constructed in \cite{EJL} from T-duality of IIA
KK-monopole in a transverse direction. Set RR field $C^{(0)}$ and
$C^{(2)}$ to zero and dilaton to a constant, the action reads:
\be\label{5braneaction} S_{NS5}=-T_{NS5}\int d^6\sigma
e^{-2\Phi}\sqrt{\mbox{det}(g+(2\pi\alpha^{'})F)} -T_{NS5}\int
d^6\sigma F\wedge C^{(4)} \ee where the tension of NS5 brane is
  \be
  T_{NS5}=\frac{1}{(2\pi)^5\alpha^{'3}g_s^2}\nonumber
  \ee
This action can also be derived from S-dual of D5-brane DBI
action. Under S-duality $c^{(1)}$ transforms to an one form gauge
field $A^{(1)}$.

In our case, the worldvolume of NS5-brane is $AdS_2\times S^4$,
located at $u=0, \theta=\theta_m=\mbox{constant}$ in coordinates
(\ref{fibre}). From the S-duality, one may expect that the
supersymmetry condition is the same as the one in D5-brane
discussed in \cite{Yamaguchi:2006D5}, namely the supersymmetry
condition requires that
 \be
 \cos\theta_m=\beta, \hspace{3ex} (1-\mu_3\tau_3)\zeta=0
 \ee
 where $\beta$ is the electric field strength on the NS5-brane.
 The similar calculation shows that
 \be
 S_{\mbox{tot}}=-\frac{4N\pi}{\sqrt{\lambda}}\frac{2N}{3\pi}\sin^3\theta_m
 \ee
 after taking into account of the boundary terms. When the string charge
 $m$ is much smaller than $N$, $\theta_m$ becomes small. Then we
 have
 \be
 S_{\mbox{tot}}\approx -\sqrt{\lambda}\frac{m}{g_s}.
 \ee
This is what we expected: the action is proportional to the
tension of $m$ overwrapping D-strings.  This is S-dual to the
D5-brane result $S_{\mbox{tot}}\approx -n \sqrt{\lambda}$.

For Wilson-'t Hooft loop, it is not easy to figure out a picture
as in D3-brane case. Now due to the existence of both F1's and
D1's, the dual bulk should have both D5 and NS5-brane. And from
the above discussion, both brane keep the same supesymmetry. But
in general D5-brane and NS5-brane are located at different
position since $\theta_k$ depends on the value of the flux. In the
case that both F1 and D1 charges are small, the total action is
just the summation of the D5 and NS5 action, namely proportional
to the summation of tensions of F1's and D1's. We are not clear if
these $F1$'s and $D1$'s form bound states and if so how to see
this property from 5-brane picture.

\section{Conclusions and Discussions}

In this paper, we discussed Wilson-'t Hooft loop in the symmetric
representation. In this case, the brane description is a D3-brane
with both electric and magnetic fluxes. Taking the boundary terms
into account properly, we obtained the expectation value of
Wilson-'t Hooft loop. The result is precisely in match with the
result from the gauge theory argument.

 In this paper, we mainly focus on Wilson-'t Hooft loops with
 symmetric representation, in which case the string picture is
 simply a single F-string with charge $n$ and a single
 D-string with charge $m$. It would be interesting to
 discuss the more general case. Let us first consider the pure
 Wilson loop in a general representation. The representation could
 be described by a Young tableau. The total number of box in the
 Young tableau is the total charges of the fundamental strings.
 Every row of the Young tableau represents a single F1 with the
 charge determined by the number of boxes in this
 row. The expectation value of Wilson loop could be calculated
 from the Hermitian matrix theory. In general, the expectation value
 is
  \be
  <\frac{1}{N}Tr\exp(k_1M)\frac{1}{N}Tr\exp(k_2M)\cdots
  \frac{1}{N}Tr\exp(k_lM)>
  \ee
  where $k_1, k_2 \cdots k_l$ are the number of boxes in each row.
  This quantity has been discussed in \cite{Akeman}.
  From our D-brane picture, the symmetric
  representations correspond to D3-brane with electric flux. One
  might naively expect that every row of the Young tableau would
  give a D3-brane with electric flux and we have an array of D3-branes
  whose number is the number of the rows to describe Wilson loop\cite{GP}.
  The configuration is still supersymmetric but the position of
  D3-branes are different due to the difference of the flux. It
  must be very interesting to check how much this D3-brane picture describe the
  physics correctly. One essential point in the D3-brane picture is that even the electric
  flux number is only one, the D3-brane description still make sense\cite{Fiol}.
  This fact indicates that we can use D3-brane configuration to describe Wilson loop in
  arbitrary representation. On the other hand we have a dual description in
  terms of D5-branes which requires that the number of F1's is
  huge. It is not clear how the two description match each other.
  For an antisymmetric description of rank $k$, we may study the problem in terms of  $k$
  D3-branes. To describe these D3-branes, we need a nonabelian DBI
  action whose form is an open question. In this sense, it would be valuable to investigate the
  correspondence between the dual pictures
  to test the nonabelian DBI action.

  It is more difficult to figure out the brane picture if including the
  D-string. As the F-strings, the D-strings could take many forms: either a
single one with charge $m$, or many ones each with winding number
one, or something between these two extremal cases. For pure
D-strings, all the above possible configurations are
supersymmetric, being S-dual to the pure F-strings. In D3-brane
picture, there are several D3-branes with different magnetic
fluxes. They are supersymmetric even though generically they don't
overlap with each other. However, if both the F1's and D1's exist,
they may form bound state. For a generic (F1, D1) configurations,
its D3/D5-brane description is unclear. Naively, one may think
that the above D3-brane picture could still make sense after
putting into the relevant magnetic flux. But from the discussion
before, we know that the supersymmetric condition is sensitive to
both the electric and magnetic fields. It is a problem on how to
distribute magnetic flux to different D3's. And even if we give an
assignation of magnetic flux to D3-brane's, the supersymmetries is
broken and there exist the open string tachyon. The tachyon
condensation lead to the bound state of (F1, D1). After tachyon
condensation, the final configuration should be independent to the
assignation of D1's. It would be interesting to understand this
process more precisely. In terms of 5-brane, the picture is more
unclear.

Another interesting problem is the dual supergravity solution of
WH loop. In \cite{Yamaguchi:2006te,Lunin:2006}, the supergravity
solution corresponding to 1/2-BPS Wilson loop is presented. The
geometry preserves $SU(1,1)\times SU(2)\times SO(5)$ isometry as
the 1/2-BPS Wilson loop itself. The geometry is governed by a
function in two dimension satisfying Laplace equation, with some
boundary conditions on $y$ axis. The boundary condition takes two
constant values, corresponding to the distribution of eigenvalues
on the real line in the matrix model. This is the fermion droplet
picture analogous to the bubbling geometry of LLM\cite{LLM}. For
1/2-BPS WH loop, the isometry is the same as Wilson loop. It would
be interesting to see what the  bubbling geometry is.

In this paper, the expectation value of 1/2 BPS Wilson-'t Hooft
loop is from the bulk D3-brane picture. It would be nice to have a
direct perturbative calculation in the SYM.

\acknowledgments{ The work was supported by NSFC Grant No.
10405028,10535060 and the Key Grant Project of Chinese Ministry of
Education (NO. 305001)}

\appendix
\section{Spinors in $AdS_2\times S^2 \times S^4$ fibration}

\label{app1}

In IIB supergravity, the supersymmetry is determined by the number
of covariant constant spinor. In this paper, we focused on
$AdS_5\times S^5$ background, where only the metric and the RR
5-form $G_5$ are nonvanishing. The supersymmetry transformation of
dilatino is trivial and the invariance of the gravitino under
supersymmetry transformation can be written as
\begin{align}
 &\delta \psi_{M}=\nabla_{M}\eta+\frac{i}{2}\sla{G}_5\tau_2 \Gamma_{M}\eta.
\end{align}
 The parameter of
supersymmetry is a doublet of Majorana-Weyl spinors:
$\eta=(\eta_1,\eta_2)$. We use the Pauli matrices $\tau_j$ to
rotate this
doublet:$(\tau\eta)_\alpha=\tau_{\alpha\beta}\eta_\beta$.

 Since we
have written $AdS_5\times S^5$ metric in forms $AdS_2\times S^2
\times S^4$ fibered on $\mathbb{R}^2$, it's convenient to write
10-dimensional Clifford algebra in tensor product form. Let
$\Gamma^\mu$ denote the Gamma matrices in ten dimension. They can
be written as the following tensor product
form\cite{Yamaguchi:2006te,Yamaguchi:2006D5}
\begin{align}
 &\Gamma^0=\sigmac^{0}\otimes \sigc \otimes 1 \otimes 1,
 &&\Gamma^1=\sigmac^{1}\otimes \sigc \otimes 1 \otimes 1,\nn\\
 &\Gamma^2=1\otimes \sigma_1\otimes 1 \otimes 1,
 &&\Gamma^3=1\otimes \sigma_2\otimes 1 \otimes 1,\nn\\
 &\Gamma^4=\sigmac^{3}\otimes \sigc \otimes \sigmah_4 \otimes 1,
 &&\Gamma^5=\sigmac^{3}\otimes \sigc \otimes \sigmah_5 \otimes 1,\nn\\
 &\Gamma^a=\sigmac^{3}\otimes \sigc \otimes \sigmah_6 \otimes \gamma^{a},
&&\qquad (a=6,7,8,9),
\end{align}
where $(\sigmac_1,\sigmac_2,\sigmac_3)$,
$(\sigma_1,\sigma_2,\sigc)$ and $(\sigmah_4,\sigmah_5,\sigmah_6)$
are three sets of the Pauli matrices. And
$\sigmac^{0}:=i\sigmac_2$. $\gamma^{a}$'s $(a=6,7,8,9)$ are gamma
matrices of Euclidean 4 dimensions.

 A 10-dimensional spinor $\eta$
can be decomposed as
\begin{align}
 \eta=\sum_{a,b,c}\chic^{I}_{a}\otimes\epsilon_{abcIJK}\otimes\chih^{J}_{b}
\otimes \chi^{K}_{c}.
\end{align}
Each $\epsilon_{abcIJK}$ is a pair of 2-dimensional spinors with
$\sigma_1,\sigma_2,\sigc,\tau_1,\tau_2,\tau_3$ acting on it. The
spinors satisfy the following relations
\begin{align}
 &\nablac_p \chic^{I}_{a}=\frac i2 a \sigmac_{p}\chic^{I}_{-a},\qquad
 \sigmac_{3}\chic^{I}_{a}=a\chic^{I}_{a},\qquad
  (p=0,1,\qquad a=\pm 1,\qquad I=1,2),\\
  &\nablac_p \chih^{J}_{b}=\frac 12 b \sigmah_{p}\chih^{J}_{-b},\qquad
 \sigmah_{6}\chih^{J}_{b}=b\chih^{J}_{b},\qquad
 (p=4,5,\qquad b=\pm 1,\qquad J=1,2),\\
  &\nablac_p \chi^{K}_{c}=\frac 12 c \gamma_{p}\chi^{K}_{-c},\qquad
 \gamma_{6789}\chi^{K}_{c}=c\chi^{K}_{c},\qquad
 (p=6,\dots,9,\qquad c=\pm 1,\qquad K=1,2,3,4).
\end{align}
where $\nablac$ is the covariant derivative of the Levi-Civita
connection of the unit $AdS_2$, $S^2$ or $S^4$. By expanding the
10-dimensional spinor pair $\eta$ by the above Killing spinors we
reduce the  problem to 2-dimensions .

With these notations, the supersymmetry condition in IIB theory
yields $\epsilon$ of the form
\begin{align}
\epsilon=\exp(\frac{1}{2}\mu_3\nu_3\tau_2
u-\frac{i}{2}\lambda_3\tau_2\theta)\zeta.
\end{align}
where $\zeta$ a covariant constant spinor, constrained by the
following conditions
\begin{align}
\mu_2\lambda_3\tau_2\Gamma^2\zeta=\zeta, \quad
\nu_1\lambda_3\Gamma^2\zeta=\zeta,\quad \lambda_1\Gamma^3\zeta=\zeta,\quad
\Gamma_{23}\zeta=-i\mu_3\nu_3\lambda_3\zeta.
\end{align}


\begin{thebibliography}{99}
\bibitem{Maldacena:1997}
J.M.Maldacena, {\it ``The large $N$ limit of superconformal field
theories and supergravity,''}Adv.\ Theor.\ Math.\ Phys.\ {\bf 2},
231 (1998) [hep-th/9711200].
\bibitem{AGMOO}
O. Aharony, S.S. Gubser, J. Maldacena, H. Ooguri, Y. Oz, {\it
``Large N Field Theories, String Theory and Gravity,''}  Phys.
Rept.
 {\bf  323}183 (2000), [\href{http://arXiv.org/abs/hep-th/9905111}{{\tt
  hep-th/9905111}}].

\bibitem{Rey:1998ik}
S.-J. Rey and J.-T. Yee, {\it ``Macroscopic strings as heavy
quarks in large
  {N} gauge theory and anti-de {S}itter supergravity,''}  Eur. Phys. J. {\bf
  C22} (2001) 379--394, [\href{http://arXiv.org/abs/hep-th/9803001}{{\tt
  hep-th/9803001}}].

\bibitem{Maldacena:1998im}
J.~M. Maldacena, {\it ``{W}ilson loops in large {N} field
theories,''}  Phys.
  Rev. Lett. {\bf 80} (1998) 4859--4862,
  [\href{http://arXiv.org/abs/hep-th/9803002}{{\tt hep-th/9803002}}].


  \bibitem{polya1}
A. Polyakov, {\it ``The wall of the cave,''} Int.J.Mod.Phys. {\bf
A14} (1999) 645-658 ,
  [\href{http://arXiv.org/abs/hep-th/9809057}{{\tt hep-th/9809057}}].

  \bibitem{Zarembo1}
K. Zarembo, {\it ``Supersymmetric Wilson loops,''} Nucl.Phys. {\bf
B643} (2002) 157-171 ,
  [\href{http://arXiv.org/abs/hep-th/hep-th/0205160}{{\tt hep-th/hep-th/0205160}}].


  \bibitem{DGGM}
A. Dymarsky, S. Gubser, Z. Guralnik, J. Maldacena, {\it
``Calibrated Surfaces and Supersymmetric Wilson Loops ,''}
  [\href{http://arXiv.org/abs/hep-th/hep-th/0604058}{{\tt hep-th/hep-th/0604058}}].


  \bibitem{Drukker:1999zq}
N.~Drukker, D.~J. Gross and H.~Ooguri, {\it ``{W}ilson loops and
minimal
  surfaces,''}  Phys. Rev. {\bf D60} (1999) 125006,
  [\href{http://arXiv.org/abs/hep-th/9904191}{{\tt hep-th/9904191}}].


  \bibitem{Erickson:2000af}
J.~K. Erickson, G.~W. Semenoff and K.~Zarembo, {\it ``Wilson loops
in {N = 4}
  supersymmetric {Y}ang-{M}ills theory,''}  Nucl. Phys. {\bf B582} (2000)
  155--175, [\href{http://arXiv.org/abs/hep-th/0003055}{{\tt hep-th/0003055}}].

\bibitem{Drukker:2000rr}
N.~Drukker and D.~J. Gross, {\it ``An exact prediction of {N = 4}
{SUSYM}
  theory for string theory,''}  J. Math. Phys. {\bf 42} (2001) 2896--2914,
  [\href{http://arXiv.org/abs/hep-th/0010274}{{\tt hep-th/0010274}}].
  \bibitem{Fiol}
N.~Drukker and B.~Fiol, {\it ``All-genus calculation of {W}ilson
loops using
  {D}-branes,''}  JHEP {\bf 02} (2005) 010,
  [\href{http://arXiv.org/abs/hep-th/0501109}{{\tt hep-th/0501109}}].

 \bibitem{Hartnoll:2006hr}
S.~A. Hartnoll and S.~{Prem Kumar}, {\it ``Multiply wound
{P}olyakov loops at
  strong coupling,''}  \href{http://arXiv.org/abs/hep-th/0603190}{{\tt
  hep-th/0603190}}.

  \bibitem{GP}
Jaume Gomis, Filippo Passerini {\it ``Holographic Wilson Loops,''}
 [\href{http://arXiv.org/abs/hep-th/0604007}{{\tt
  hep-th/0604007}}].

\bibitem{Akeman}G. Akemann and P.H. Damgaard, {\it ``Wilson loops in
N=4 supersymmetric Yang-Mills theory from random matrix theory"},
Phys. Lett. {\bf B513}(2001)179, Erratum-ibid. {\bf
B524}(2002)400, [\href{http://arXiv.org/abs/hep-th/0101225}{{\tt
hep-th/0101225}}].


  \bibitem{OS}
Kazumi Okuyama, Gordon W. Semenoff, {\it ``Wilson Loops in N=4 SYM
and Fermion Droplets,''}
  [\href{http://arXiv.org/abs/hep-th/0604209}{{\tt hep-th/0604209}}].

\bibitem{OGS}
Sean A. Hartnoll, S. Prem Kumar,
 {\it ``Higher rank Wilson loops from a matrix model,''}
  [\href{http://arXiv.org/abs/hep-th/0605027}{{\tt hep-th/0605027}}].

\bibitem{Myers:1999}
R.C. Myers, {\it ``Dielectric-Branes,''}  JHEP 9912 (1999) 022,
  [\href{http://arXiv.org/abs/hep-th/9910053}{{\tt hep-th/9910053}}].

\bibitem{Pawelczyk:2000hy}
J.~Pawelczyk and S.-J. Rey, {\it ``{R}amond-{R}amond flux
stabilization of
  {D}-branes,''}  Phys. Lett. {\bf B493} (2000) 395--401,
  [\href{http://arXiv.org/abs/hep-th/0007154}{{\tt hep-th/0007154}}].

\bibitem{Rodriguez2006}D. Rodriguez-Gomez, {\it ``Computing Wilson
lines with dielectric branes"},
[\href{http://arXiv.org/abs/hep-th/0604031}{{\tt
hep-th/0604031}}].

\bibitem{Grisaru:2000zn}
M.~T. Grisaru, R.~C. Myers and O.~Tafjord, {\it ``{SUSY} and
{G}oliath,''}
  JHEP {\bf 08} (2000) 040, [\href{http://arXiv.org/abs/hep-th/0008015}{{\tt
  hep-th/0008015}}].

\bibitem{Hashimoto:2000zp}
A.~Hashimoto, S.~Hirano and N.~Itzhaki, {\it ``Large branes in
{AdS} and their
  field theory dual,''}  JHEP {\bf 08} (2000) 051,
  [\href{http://arXiv.org/abs/hep-th/0008016}{{\tt hep-th/0008016}}].

\bibitem{McGreevy:2000cw}
J.~McGreevy, L.~Susskind and N.~Toumbas, {\it ``Invasion of the
giant gravitons
  from anti-de {S}itter space,''}  JHEP {\bf 06} (2000) 008,
  [\href{http://arXiv.org/abs/hep-th/0003075}{{\tt hep-th/0003075}}].

\bibitem{WITTENboun}
Witten, Edward,
 {\it ``Bound States Of Strings And p-Branes,''}  Nucl.Phys. {\bf B460} (1996) 335-350 ,
  [\href{http://arXiv.org/abs/hep-th/9510135}{{\tt hep-th/9510135}}].

\bibitem{Kapustin2005}A. Kapustin, {\it ``Wilson-'t Hooft
operators in four-dimensional gauge theories and S-duality"},
[\href{http://arXiv.org/abs/hep-th/0501015}{{\tt
hep-th/0501015}}].



\bibitem{Yamaguchi:2006te}
S.~Yamaguchi, {\it ``Bubbling geometries for half {BPS} {W}ilson
lines,''}
  \href{http://arXiv.org/abs/hep-th/0601089}{{\tt hep-th/0601089}}.

\bibitem{Lunin:2006}
Oleg Lunin, {\it ``On gravitational description of Wilson
lines,''}
  \href{http://arXiv.org/abs/hep-th/0604133}{{\tt hep-th/0604133}}.

\bibitem{Yamaguchi:2006D5}
S.~Yamaguchi, {\it ``Wilson Loops of Anti-symmetric Representation
and D5-branes,''} JHEP 0605 (2006) 037
  \href{http://arXiv.org/abs/hep-th/0603208}{{\tt hep-th/0603208}}.

\bibitem{BGK}
Massimo Bianchi, Michael B. Green, Stefano Kovacs, {\it ``
Instanton corrections to circular Wilson loops in N=4
Supersymmetric Yang-Mills,''} JHEP 0204 (2002) 040
  \href{http://arXiv.org/abs/hep-th/0202003}{{\tt hep-th/0202003}}.




\bibitem{LLM}
H.~Lin, O.~Lunin and J.~Maldacena, {\it ``Bubbling {AdS} space and
1/2 {BPS}
  geometries,''}  JHEP {\bf 10} (2004) 025,
  [\href{http://arXiv.org/abs/hep-th/0409174}{{\tt hep-th/0409174}}].


\bibitem{Callan:1997}
C.~G. Callan, Jr. and J.~M. Maldacena, {\it ``Brane dynamics from
the
  {B}orn-{I}nfeld action,''}  Nucl. Phys. {\bf B513} (1998) 198--212,
  [\href{http://arXiv.org/abs/hep-th/9708147}{{\tt hep-th/9708147}}].


\bibitem{Bergshoeff:1996}
E.~Bergshoeff and P.~K. Townsend, {\it ``Super {D}-branes,''}
Nucl. Phys. {\bf
  B490} (1997) 145--162, [\href{http://arXiv.org/abs/hep-th/9611173}{{\tt
  hep-th/9611173}}].



\bibitem{Bergshoeff:1997}
E.~Bergshoeff, R.~Kallosh, T.~Ortin and G.~Papadopoulos, {\it
``kappa-symmetry,
  supersymmetry and intersecting branes,''}  Nucl. Phys. {\bf B502} (1997)
  149--169, [\href{http://arXiv.org/abs/hep-th/9705040}{{\tt hep-th/9705040}}].

\bibitem{Skenderis:2002vf}
K.~Skenderis and M.~Taylor, {\it ``Branes in {AdS} and pp-wave
spacetimes,''}
  JHEP {\bf 06} (2002) 025, [\href{http://arXiv.org/abs/hep-th/0204054}{{\tt
  hep-th/0204054}}].

\bibitem{Imamura:1998gk}
Y.~Imamura, {\it ``Supersymmetries and {BPS} configurations on
anti-de {S}itter
  space,''}  Nucl. Phys. {\bf B537} (1999) 184--202,
  [\href{http://arXiv.org/abs/hep-th/9807179}{{\tt hep-th/9807179}}].



\bibitem{EJL}
Eduardo Eyras, Bert Janssen, Yolanda Lozano, {\it ``5-branes,
KK-monopoles and T-duality,''} Nucl.Phys. {\bf B531} (1998)
275-301,
  [\href{http://arXiv.org/abs/hep-th/9806169}{{\tt hep-th/9806169}}].



\end{thebibliography}
\end{document}